\documentclass[aps,prb,reprint,amsfont,amsmath,amssymb,floatfix]{revtex4-2}
\usepackage{graphicx,bm,xcolor}
\usepackage[normalem]{ulem}
\usepackage{hyperref}
\hypersetup{colorlinks=true,allcolors=blue}

\begin{document}

\title{Hall effect in slip flow of two-dimensional electron fluid}
\author{A.~A.~Grigorev}
\affiliation{Ioffe Institute, St.~Petersburg 194021, Russia}
\author{ A.~N.~Afanasiev}
\email{afanasiev.an@mail.ru}
\affiliation{Ioffe Institute, St.~Petersburg 194021, Russia}

\begin{abstract}
Hall effect in high-mobility 2D mesoscopic samples with hydrodynamic electron transport is related to manifestation of non-dissipative Hall viscosity at classical magnetic fields. However, the latter can be obscured by the particular geometry of the electronic flow through constriction and boundary effects. In this work we consider the low-field Hall resistivity of the narrow channel with hydrodynamic electron transport. Using kinetic theory we show that the hydrodynamic Hall viscosity contribution is accompanied by the comparable ballistic contribution associated with the formation of Knudsen layers in the slip flow of 2D electron fluid. The studied ballistic correction is present at any specularity of the boundary reflection and lowers the magnitude of the total negative size-dependent correction to the bulk Hall resistivity. The obtained analytic expression for the total Hall resistivity can be used to estimate the boundary reflection coefficient in high-mobility structures.
\end{abstract}

\maketitle

%%%%%%%%%%%%%%%%%%%%%%%%%%%%%%%%%%%%%%%%%%%%%%%%%%%%%%%%%%%%%%%%%%%%%%%%%%%%%
%%%%%%%%%%%%%%%%%%%%%%%%%%%%%%%%%%%%%%%%%%%%%%%%%%%%%%%%%%%%%%%%%%%%%%%%%%%%%

{\em 1. Introduction.}
Hydrodynamic regime of electron transport occurs in mesoscopic structures when the processes of momentum-conserving electron-electron scattering dominate over momentum-relaxing scattering by disorder, phonons and the edges of the sample. In these conditions the electronic flow becomes collective and the transport is determined by the shear viscosity of electron fluid~\cite{Gurzhi1968,Polini2020,Narozhny2022, Varnavides2023}. Various signatures of the viscous flows of electron fluid were observed in modern high-mobility 2D structures of graphene~\cite{Bandurin2016, Crossno2016, Sulpizio2019, Ku2020}, GaAs quantum wells~\cite{Gusev2018AIP, Alekseev2016, Gusev2018, Wang2022} and 2D metals~\cite{Moll2016}.

Hall effect in the samples with hydrodynamic regime of electron transport is marked by unusual contribution associated with the non-dissipative Hall viscosity~\cite{Steinberg1958, Kaufman1960, Avron1995, Avron1998, Alekseev2016} at classical magnetic fields. It leads~\cite{Scaffidi2017} to the negative size-dependent correction $\Delta \rho_{xy}^h=\rho_{xy}^h-\rho_{xy}^{\rm bulk}$ to the bulk Hall coefficient $\rho_{xy}^{\rm bulk}=-B/N_0 e c$ ($N_0$ is the resident carrier density) which determines the temperature evolution of the Hall resistivity for constrained geometries~\cite{Gusev2018, Berdyugin2019, Raichev2020, Wang2023}. The effect is maximal at low $B$~\cite{Scaffidi2017} due to suppression of shear viscosities of electron fluid~\cite{Steinberg1958, Kaufman1960, Alekseev2016} by magnetic field. Owing to the viscous nature of correction, spatial gradients of the flow velocity play decisive role which induces sensitivity of the effect to the particular geometry of the sample and boundary scattering. Therefore, considerable theoretical and experimental efforts have been made to disentangle the effect of the Hall viscosity from the sample geometry~\cite{Pellegrino2017, Holder2019, Berdyugin2019}.

For the most common long and narrow channel geometry both experimental~\cite{Gusev2018, Raichev2020, Wang2023} and numerical~\cite{Scaffidi2017} studies show that the low-field hydrodynamic Hall viscosity contribution for 2D Poiseuille flow $\Delta \rho_{xy}^{\rm h}/\rho_{xy}^{\rm bulk}=-6\,{\rm Kn}^2$ systematically underestimates the total transverse resistivity. Here ${\rm Kn}=l_{ee}/W$ is the Knudsen number which describes the degree of ballisticity of the flow, $l_{ee}$ is the viscosity electron-electron mean free path~\cite{Principi2016,AlekseevDmitriev2020} and $W$ is the channel width. This discrepancy is caused~\cite{Afanasiev2022} by the influence of the near-edge Knudsen layers~\cite{Raichev2022, Raichev2022cond, Raichev2023, Afanasiev2025}, which appear within the slip corrections (next order by ${\rm Kn}\ll 1$) to the Poiseuille flow and represent ballistic contribution to the purely hydrodynamic transport~\cite{Afanasiev2025}.

In this work we focus on ballistic corrections to the low-field Hall resistivity of the narrow channel with 2D electron fluid in weak magnetic field. In the framework of the Boltzman transport equation in relaxation time approximation supplemented by the partially diffusive/specular boundary scattering model we perform analytic calculation of $\Delta \rho_{xy}/\rho_{xy}^{\rm bulk}$. We show that the effect of Hall viscosity in transverse resistivity is mixed with the comparable ballistic contribution due to Knudsen layers of the slip part of the flow. This result indicates that the role of Knudsen layers in hydrodynamic transport is promoted by magnetic field. We determine the dependence of ballistic contribution to $\rho_{xy}$ on the boundary reflection coefficient and discuss how the boundary scattering in the sample can be characterized via the measurements of the low-field Hall resistivity.

%%%%%%%%%%%%%%%%%%%%%%%%%%%%%%%%%%%%%%%%%%%%%%%%%%%%%%%%%%%%%%%%%%%%%%%%%%%%%
%%%%%%%%%%%%%%%%%%%%%%%%%%%%%%%%%%%%%%%%%%%%%%%%%%%%%%%%%%%%%%%%%%%%%%%%%%%%%

{\em 2. Kinetic equation for the narrow channel geometry in magnetic field.}
Semiclassical electron transport in mesoscopic structures is conventionally described by the Boltzman kinetic equation. For stationary 2D flows driven by the constant electric field $\textbf{E}_0$ in the presence of perpendicular magnetic field $\textbf{B}$, its linearized form is
\begin{equation}
	\label{Eq:KE_general}
	\left[\textbf{v}_{\bf p}\bm{\nabla}+\omega_c\partial_{\theta}\right]\delta n - e\textbf{E}\textbf{v}_{\bf p}n'_0={\rm St}[\delta n]\,,
\end{equation}
where $\textbf{v}_{\bf p}=\textbf{p}/m$ is the quasiparticle velocity, $\omega_c=eB/mc$ is the cyclotron frequency, angle $\theta$ indicates the direction of electron's momentum $\textbf{p}$, $\textbf{E}(\textbf{r})=\textbf{E}_0+\textbf{E}_H(\textbf{r})$ is the total electric field which takes into account the Hall component $\textbf{E}_{H}(\textbf{r})$. Here $\delta n_{\textbf{p}}(\textbf{r})=n_{\textbf{p}}(\textbf{r})-n_0(\epsilon_{\textbf{p}})$ stands for the deviation of the nonequilibrium distribution from the global equilibrium one $n_0(\epsilon_{\textbf{p}})$ described by the Fermi-Dirac distribution, where $\epsilon_{\textbf{p}}=p^2/2m$ is the energy of non-interacting quasiparticles. To describe the effect of the dominant momentum-conserving electron-electron scattering on the distribution function in Eq.~(\ref{Eq:KE_general}), we use the relaxation time approximation for the linearized collision integral ${\rm St}[\delta n]=-\tau_{ee}^{-1}(\delta n + \textbf{p}\textbf{V} n'_0)$. In this simplified form of ${\rm St}[\delta n]$ deviations from the local equilibrium $n_{\rm loc}=-\textbf{p}\textbf{V}n'_0$ characterized by the average flow velocity $\textbf{V}(\textbf{r})=2\sum_{\textbf{p}} \textbf{v}_{\textbf{p}} \delta n_{\textbf{p}}(\textbf{r})$ decay with the single viscosity lifetime $\tau_{ee}$~\cite{Principi2016, AlekseevDmitriev2020}. In general, local equilibrium distribution is also determined by the nonequilibrium part of the chemical potential $\delta \mu(\textbf{r})=2\sum_{\textbf p}\delta n_{\textbf{p}}(\textbf{r})$. Here we focus on the slow flow regime at low Mach numbers ${\rm M}=V/s\ll 1$, where $s$ is the speed of sound~\cite{Speed_Note} and thus neglect the compressibility of electron fluid $\delta\mu(\textbf{r})\approx 0$.

Hydrodynamic transport in high-quality samples of GaAs quantum wells and graphene typically occurs at moderate temperatures $T$ and resident carrier densities $N_0$ when the 2D electron system is degenerate $T\ll E_F$ ($E_F$ is the Fermi energy). Therefore, we focus on the low-temperature limit when $n_0(\epsilon_{\textbf{p}})=\Theta(E_F-\epsilon_{\textbf{p}})$ and parametrize weakly nonequilibrium distribution of carries in terms of the small rigid body deformation of the Fermi surface $\Phi(\textbf{r},\theta)\ll E_F$~\cite{Call_Note}:
\begin{equation}
	\label{Eq:Deformed_FS}
	n_{\textbf{p}}(\textbf{r})=n_0(\epsilon_{\textbf{p}}-\Phi(\textbf{r},\theta))\approx n_0(\epsilon_{\textbf{p}}) - n'_0(\epsilon_{\textbf{p}}) \Phi(\textbf{r},\theta)\,.
\end{equation}

For the long and narrow channel, electron distribution and flow characteristics are uniform along it ($Oy$ axis collinear to the applied field $\textbf{E}_0$) and vary in the transverse direction ($Ox$ axis collinear to the Hall field $\textbf{E}_H$). Since the boundaries of the channel are current-impenetrable and ${\rm div}\textbf{V}=0$ due to incompressibility of the fluid, the flow along the channel is one-dimensional with the velocity field components $V_y(x)=V(x)$ and $V_x\equiv 0$. In magnetic field, it is convenient to describe the flow by the auxiliary distribution with $\tilde{\Phi}(x,\theta)=\Phi(x,\theta)-e\phi_{H}(x)$~\cite{AlekseevSemina2019, Raichev2020, Afanasiev2021, Afanasiev2022, Raichev2023}, where $\phi_H(x)$ is the electrostatic potential of the Hall field $E_H(x)=-\partial_x \phi_H(x)$. Finally, after substitution of Eq.~(\ref{Eq:Deformed_FS}) into Eq.~(\ref{Eq:KE_general}) kinetic equation for the transformed deformation of the Fermi surface $\tilde{\Phi}(x,\theta)$ reads
\begin{gather}
	\label{Eq:KE_Phi}
	\tilde{\Phi}=\tilde{\Phi}_{\rm loc}-e E_0 l_{ee}\sin\theta-\cos\theta l_{ee}\partial_x \tilde \Phi -\omega_c\tau_{ee} \partial_{\theta}\tilde{\Phi}\,,\\
	\label{Eq:Phi_loc}
	\tilde{\Phi}_{\rm loc}(x,\theta)=-e\phi_H(x)+p_F V(x) \sin\theta\,,
\end{gather}
where $l_{ee}=v_F \tau_{ee}$ and $\theta=\widehat{\textbf{v}_F \textbf{e}_{x}} \in [-\pi/2,3\pi/2]$. Instead of the additional force term like in Eq.~(\ref{Eq:KE_general}), the Hall field is present in Eq.~(\ref{Eq:KE_Phi}) in the form of nonequilibrium electrochemical potential $\delta\mu(x)-e\phi_{H}(x)$ (at nonzero $\delta\mu$) in the modified local equlibrium distribution~(\ref{Eq:Phi_loc}). Kinetic equation~(\ref{Eq:KE_Phi}) takes the closed form by introducing connection between the macroscopic flow characteristics and the angular harmonics of the distribution function $-e\phi_H(x)=\langle\tilde{\Phi}(x,\theta)\rangle$, $V(x)=2\langle\tilde{\Phi}(x,\theta)\sin\theta\rangle/p_F$ where $\langle\dots\rangle=\int_{-\pi/2}^{3\pi/2}d\theta/2\pi$ denotes averaging over directions of electron's velocity.

At the rough current-impenetrable boundaries of the channel (at $x=\pm W/2$) the incident electrons undergo uncorrelated scattering and the corresponding boundary conditions for $\tilde{\Phi}(x,\theta)$ have the form~\cite{deJong1995, Raichev2020, Raichev2023}
\begin{multline}
	\label{Eq:BC}
	\tilde{\Phi}_{\pm}(\mp W/2,\theta)=r\tilde{\Phi}_{\mp}(\mp W/2,\pi-\theta)\mp\\
	\mp\frac{1-r}{2}\langle\cos\theta' \tilde{\Phi}_{\mp}\left(\mp W/2,\theta'\right)\rangle_{\mp}\,.
\end{multline}
Here $\pm$ denote right $\theta_{+}\in[-\pi/2,\pi/2]$ and left $\theta_{-}\in[\pi/2,3\pi/2]$ traveling electrons (with respect to $Ox$) and $\langle\dots\rangle_{\pm}=\int_{\pm}d\theta/2\pi$. Eq.~(\ref{Eq:BC}) connects the distributions of incident (on the right-hand side) and reflected (on the left-hand side) electrons in result of partly specular and partly diffusive scattering by the channel walls. The degree of specularity is described by the reflection coefficient $0\leq r \leq 1$. At $r=1$ the boundary scattering is purely specular $\tilde{\Phi}_{\pm}(\mp W/2,\theta)=\tilde{\Phi}_{\mp}(\mp W/2,\pi-\theta)$ and for $r=0$ it is purely diffusive $\tilde{\Phi}_{\pm}(\mp W/2,\theta)={\rm const}$. In this work we consider the model of angle independent $r$. Note that Eq.~(\ref{Eq:BC}) ensures $V_y(\pm W/2)=0$.

In weak magnetic fields $\omega_c\tau_{ee}=l_{ee}/R_c\ll 1$ ($R_c=v_F/\omega_c$ is the cyclotron radius) electron trajectories between the frequent interparticle collisions are only slightly bent and the distribution function is analytic in $\omega_c\tau_{ee}$ across the whole channel. The effect of magnetic field on electronic transport in this regime can be studied perturbatively and the distribution function can be represented in the form of expansion $\tilde{\Phi}(x,\theta)=\sum_n\tilde{\Phi}^{(n)}(x,\theta)$ where $\tilde{\Phi}^{(n)}(x,\theta)\sim(\omega_c\tau_{ee})^n$. The Hall effect is governed by the first order correction $\tilde{\Phi}^{(1)}(x,\theta)$ for which the kinetic equation~(\ref{Eq:KE_Phi}) is reduced to
\begin{equation}
	\label{Eq:KE_Phi_1}
	\tilde{\Phi}^{(1)}=-e\phi_H-\cos\theta l_{ee}\partial_x\tilde{\Phi}^{(1)}-\omega_c\tau_{ee}\partial_{\theta}\Phi^{(0)}\,,
\end{equation}
while the boundary conditions for $\tilde{\Phi}^{(1)}(x,\theta)$ are given by Eq.~(\ref{Eq:BC}). We note here that $\phi_{H}(x)$ is the odd function of $B$ while $V(x)$ is even and thus the latter remains unchanged in the first order by $\omega_c\tau_{ee}\ll1$.

The inhomogeneous term in~(\ref{Eq:KE_Phi_1}) is determined by $\Phi^{(0)}(x,\theta)$ which describes the zero field transport in the narrow channel considered previously in Ref.~\cite{Afanasiev2025}. Here we briefly summarize the main results of~\cite{Afanasiev2025}. In the leading order by small Knudsen number ${\rm Kn}\ll1$ stationary electronic transport in hydrodynamic narrow channel (i.e. when $l_{ee} \ll W \ll l_p$ where $l_p$ is the momentum-relaxing mean free path) is described by 2D Poiseuille flow~\cite{Alekseev2016}. However, since the modern experimentally relevant structures are characterized by small but finite ${\rm Kn}$~\cite{GrigorevAfanasiev2025}, one needs to take into account the first order corrections by ${\rm Kn}$ to $\Phi^{(0)}(x,\theta)$ which describe the slip flow of electron fluid. The corresponding distribution function of carries consists of two parts $\Phi^{(0)}_{\pm}(x,\theta)=\Phi^{(0)}_h(x,\theta)+\Phi^{(0)}_{K\pm}(x,\theta)$. The former describes the flow of hydrodynamic electrons in the whole channel under external field $\textbf{E}_0$ which is controlled by electron-electron scattering. The flow velocity profile of this hydrodynamic part is inhomogeneous on the characteristic scale of the channel width $W$ and acquires finite slip at the boundaries. The hydrodynamic part of the distribution $\Phi^{(0)}_h(x,\theta)$ slowly varies on the microscopic scale of $l_{ee}$ and is analytic in ${\rm Kn}$. The latter term $\Phi^{(0)}_{K\pm}(x,\theta)$ describes the semiballistic flow in the near-edge Knudsen layers with the width of the order of $l_{ee}$. Their origin is closely related to the boundary scattering of the incident hydrodynamic electrons which introduces local non-hydrodynamic perturbation of the overall distribution. Boundary distribution of the reflected electrons determined by the kinematic boundary conditions~(\ref{Eq:BC}) relaxes to its hydrodynamic form due to electron-electron collisions towards the center of the channel leading to the high degree of ballisticity of the flow in the Knudsen layers. Thus, $\Phi^{(0)}_{K\pm}(x,\theta)$ and the corresponding flow velocity profile are non-analytic in ${\rm Kn}$. We note here that the magnitude of the Knudsen part of the distribution near the boundaries is of the order of the slip correction to $\Phi^{(0)}_{h}(x,\theta)$.

In a similar way we treat hydrodynamic and Knudsen layers contributions to $\tilde{\Phi}^{(1)}_{\pm}(x,\theta)=\tilde{\Phi}^{(1)}_h(x,\theta)+\tilde{\Phi}^{(1)}_{K\pm}(x,\theta)$ governed by Eq.~(\ref{Eq:KE_Phi_1}) to study ballistic corrections to the low-field Hall resistivity in narrow channels with hydrodynamic electron transport. Due to the qualitatively different spatial and angular structure of the hydrodynamic and Knudsen parts of the distribution, functions $\tilde{\Phi}^{(1)}_h(x,\theta)$ and $\tilde{\Phi}^{(1)}_{K\pm}(x,\theta)$ satisfy Eq.~(\ref{Eq:KE_Phi_1}) separately. Connection between them is established via the boundary conditions~(\ref{Eq:BC}) for the total distribution function.

%%%%%%%%%%%%%%%%%%%%%%%%%%%%%%%%%%%%%%%%%%%%%%%%%%%%%%%%%%%%%%%%%%%%%%%%%%%%%
%%%%%%%%%%%%%%%%%%%%%%%%%%%%%%%%%%%%%%%%%%%%%%%%%%%%%%%%%%%%%%%%%%%%%%%%%%%%%

{\em 3. Viscous contribution to the Hall effect.}
For the hydrodynamic part of the flow it is convenient to represent the first-order correction to the distribution function $\tilde{\Phi}^{(1)}_h=-(\hat{1}+\hat{D})^{-1}\left[e\phi_H^h+\omega_c\tau_{ee}\partial_{\theta}\Phi_h^{(0)}\right]$ as a formal solution of Eq.~(\ref{Eq:KE_Phi_1}), where $\hat{D}=\cos\theta l_{ee}\partial_x$, $\hat{1}$ is the unitary operator and $\Phi^{(0)}_h(x,\theta)=p_F V_h(x)\sin\theta+2 m \bar{\Pi}_{xy}^{h}(x)\sin2\theta$ is the hydrodynamic distribution at $B=0$. Here $\bar{\Pi}_{xy}^{h}(x)=-\eta\partial_x V_h(x)$ is the mass density of the viscous momentum flux and $\eta=\frac{1}{4}v_{F}l_{ee}$ is the dissipative (even) part of the kinematic shear viscosity of 2D electron gas~\cite{Alekseev2016}. Since for hydrodynamic electrons $\hat{D}\sim {\rm Kn}\,\cos\theta\ll1$, the inverse operator $(\hat{1}+\hat{D})^{-1}\approx \hat{1} + \hat{D}-\hat{D}^2$ can be approximated by the truncated Neumann series. To the leading order in Knudsen number ${\rm Kn}\ll 1$ the hydrodynamic component of $\tilde{\Phi}_{\pm}^{(1)}(x,\theta)$ is given by
\begin{equation}
	\label{Eq:Phi_h_1}
	\tilde{\Phi}^{(1)}_h(x,\theta)=-e\phi_H^h(x)+2m\bar{\Pi}_{xx}^{h}(x)\cos2\theta\,,
\end{equation}
where $\bar{\Pi}_{xx}^h(x)=-2\omega_c\tau_{ee}\bar{\Pi}_{xy}^h(x)=\eta_H \partial_x V_h(x)$ is the magnetic-field induced momentum flux controlled by the Hall (odd) viscosity $\eta_{H}=2\omega_c\tau_{ee} \eta$~\cite{Steinberg1958, Kaufman1960, Avron1995, Avron1998, Alekseev2016}. To guarantee the absence of the transverse current $V_x(x)\equiv 0$, Eq.~(\ref{Eq:Phi_h_1}) is accompanied by condition
\begin{equation}
	\label{Eq:E_H_hd}
	\partial_x \bar{\Pi}_{xx}^h+\omega_c V_h+\frac{e E_H^h}{m}=0\,,
\end{equation}
which is the $Ox$ projection of the Navier-Stokes equation for the stationary flow of 2D electron fluid in narrow channel in perpendicular magnetic field~\cite{Alekseev2016}.

Expression for the hydrodynamic part of the Hall voltage $U_H^h=-\int_{-W/2}^{W/2} E_H^h(x) dx$ immediately follows from~(\ref{Eq:E_H_hd}) and the profile of momentum flux in symmetric channel $\bar{\Pi}_{xy}^{h}(x)=-e E_0 x/m$ and has the form
\begin{equation}
	\label{Eq:U_H_hd}
	U_H^h=\rho_{xy}^{\rm bulk}I_h+\frac{\eta_H}{\eta}E_0 W\,.
\end{equation}
Eq.~(\ref{Eq:U_H_hd}) determines both kinematic and the Hall viscosity contributions to the transverse resistivity $\rho_{xy}^h=U_H^h/I_h$, where $I_h=-e N_0\int_{-W/2}^{W/2} V_h(x) dx$ is the net current. To the leading order in ${\rm Kn}$ the hydrodynamic transport through the channel is described by the 2D Poiseuille flow with parabolic velocity profile $V_P(x)$. Therefore, by approximating $I_h\approx I_P=e^2 N_0 E_0 W^3/12 m \eta$ we arrive at the well-known~\cite{Scaffidi2017, Holder2019BH, Afanasiev2022} result for the low-field Hall viscosity contribution to the transverse resistivity of the narrow channel
\begin{equation}
	\label{Eq:R_xy_h}
	\Delta\rho_{xy}^h/\rho_{xy}^{\rm bulk}=-6 {\rm Kn}^2.
\end{equation}

%%%%%%%%%%%%%%%%%%%%%%%%%%%%%%%%%%%%%%%%%%%%%%%%%%%%%%%%%%%%%%%%%%%%%%%%%%%%%
%%%%%%%%%%%%%%%%%%%%%%%%%%%%%%%%%%%%%%%%%%%%%%%%%%%%%%%%%%%%%%%%%%%%%%%%%%%%%

{\em 4. Knudsen layers contribution to the Hall effect.}
The Knudsen part of the flow and the distribution which corresponds to the near-edge layer at the one side of the channel is localized on a scale $l_{ee}$ near its boundary and thus is exponentially attenuated by a factor of $\exp(-1/{\rm Kn})$ near the opposite side~\cite{Afanasiev2025}. Therefore, at any boundary of the channel the distribution of the incident electrons is hydrodynamic with high accuracy
\begin{equation}
	\label{Eq:Phi_inc}
	\tilde{\Phi}^{(1)}_{\pm}(\pm W/2,\theta)\approx \tilde{\Phi}^{(1)}_h(\pm W/2,\theta)\,.
\end{equation}
Taking into account Eq.~(\ref{Eq:Phi_inc}), the Knudsen layers contribution to the Hall voltage can be directly obtained from the kinematic boundary conditions for the total distribution function~(\ref{Eq:BC}). By taking averages of~(\ref{Eq:BC}) and~(\ref{Eq:Phi_inc}) over velocity directions of the right and left traveling electrons we calculate the total Hall potentials $-e\phi_H^{tot}(x)=\sum_{\pm}\langle\tilde{\Phi}_{\pm}^{(1)}(x,\theta)\rangle_{\pm}$ at the boundaries $x=\pm W/2$ of the channel and obtain the total Hall voltage $U_H^{tot}=\phi_{H}^{tot}(W/2)-\phi_{H}^{tot}(-W/2)$ which has the form
\begin{gather}
	U_H^{\rm tot}=U_H^h+U_H^K\,,\\
	\label{Eq:U_H_K}
	U_H^K=-(1-r)\frac{\eta_H}{3\eta}E_0W\,,
\end{gather}
where $U_H^h$ is the hydrodynamic contribution~(\ref{Eq:U_H_hd}) and $U_H^k$ is the Khudsen layers contributions. As we see, it is also determined by the Hall viscosity.

The total Hall resistivity of the narrow channel is then given by $\rho_{xy}=(U_H^h+U_H^K)/(I_h+I_K)$, where $I_K$ is the net current in the Knudsen layers. Since $I_K\sim {\rm Kn}^2 I_P$~\cite{Afanasiev2025}, the Knudsen layers contribution to the total transverse resistivity $\rho_{xy}=\rho_{xy}^{\rm bulk}+\Delta\rho_{xy}^h+\Delta \rho_{xy}^K$ takes the form
\begin{equation}
	\label{Eq:R_xy_K}
	\Delta\rho_{xy}^{K}=-\rho_{xy}^{\rm bulk}\frac{I_K}{I_P}+\frac{U_H^K}{I_P}\,.
\end{equation}
We stress that even though the slip effects in flow of electron fluid at $B=0$ appear as higher order corrections to the Poiseuille flow in ${\rm Kn}\ll 1$, Eq.~(\ref{Eq:R_xy_K}) shows that ballistic contribution to the Hall resistivity associated with the Knudsen layers of the slip component of the flow are comparable to the Hall viscosity term at any magnitude of ${\rm Kn}$ (until ${\rm Kn}<1$), as discussed in~\cite{Afanasiev2022}. Finally, using the explicit expression for the net current in Knudsen layers $I_K=3 I_P {\rm Kn}^2 (1+r)[1-\xi(r)/\xi_0(r)]$~\cite{Afanasiev2025} where $\xi_0(r)=\frac{3\pi}{16}\frac{1+r}{1-r}l_{ee}$ is the Maxwell slip length~\cite{Kiselev2019, Raichev2022} and $\xi(r)\approx1.187\xi_0(r)$ is the (exact) kinematic slip length~\cite{Afanasiev2025}, the total correction to the bulk Hall resistivity for the narrow channel in weak magnetic field is given by
\begin{gather}
	\Delta\rho_{xy}/\rho_{xy}^{\rm bulk}=-6\beta_H(r) {\rm Kn}^2\,,\\
	\beta_H(r)=\frac{7}{6}+\frac{5 r}{6}-\frac{(1+r)\xi(r)}{2\xi_0(r)}\,.
\end{gather}
The values of the coefficient $\beta_h(r)$ are below its purely viscous limit $\beta_H(r)\to 1$ for any boundary reflection coefficient $0\leq r\leq 1$ in agreement with~\cite{Scaffidi2017, Gusev2018, Wang2023}. Therefore, Knudsen layers correction~(\ref{Eq:R_xy_K}) is present at any $r$ but the partial terms behave differently with the degree of boundary specularity. In particular, the second term in~(\ref{Eq:R_xy_K}) associated with the boundary relaxation of the diagonal component of the shear stress $\bar{\Pi}_{xx}(x)=\langle\tilde{\Phi}(x,\theta)\cos2\theta\rangle/m$ vanishes for the purely specular reflection mechanism $r=1$, while the first one reminiscent the kinematic resistivity of ballistic channels~\cite{AlekseevSemina2019} is not. The reason beyond this asymmetry is that formation of the Knudsen layers at $B=0$ is related to the boundary relaxation of the non-diagonal shear stress $\bar{\Pi}_{xy}(x)=\langle\tilde{\Phi}(x,\theta)\sin2\theta\rangle/m$ present at any $r$ while the diagonal component $\bar{\Pi}_{xx}$ is conserved in the purely specular reflection.

%%%%%%%%%%%%%%%%%%%%%%%%%%%%%%%%%%%%%%%%%%%%%%%%%%%%%%%%%%%%%%%%%%%%%%%%%%%%%
%%%%%%%%%%%%%%%%%%%%%%%%%%%%%%%%%%%%%%%%%%%%%%%%%%%%%%%%%%%%%%%%%%%%%%%%%%%%%

{\em 5. Conclusion.}
In conclusion, we have developed kinetic theory of the low-field Hall resistivity in narrow channels with hydrodynamic transport regime and shown that ballistic correction to $\rho_{xy}^{\rm bulk}$ associated with Knudsen layers mixes with the Hall viscosity contribution at any specularity of the boundaries. The obtained analytical expression for $\Delta\rho_{xy}$ can be used to estimate the reflection coefficient of the boundaries and thus characterize the boundary scattering in mesoscopic high-mobility samples.

%%%%%%%%%%%%%%%%%%%%%%%%%%%%%%%%%%%%%%%%%%%%%%%%%%%%%%%%%%%%%%%%%%%%%%%%%%%%%
%%%%%%%%%%%%%%%%%%%%%%%%%%%%%%%%%%%%%%%%%%%%%%%%%%%%%%%%%%%%%%%%%%%%%%%%%%%%%

{\em 6. Acknowledgments.}
The authors are grateful to P. S. Alekseev for valuable discussions. The work was supported by the Foundation for the Advancement of Theoretical Physics and Mathematics ``BASIS'' (Grant No. 23-1-2-25).

%\begin{acknowledgments}
%\end{acknowledgments}

\bibliography{KnHE_bib}

\begin{thebibliography}{38}
\bibitem [{Spe()}]{Speed_Note}%
  \BibitemOpen
  \href@noop {} {\ }\bibinfo {note} {In degenerate 2D electron gas the speed of
  sound is $s=v_F/\sqrt{2}$.}\BibitemShut {Stop}%
\end{thebibliography}

\end{document}